\documentclass[aps,prb, twocolumn, letterpaper]{revtex4}
\usepackage{graphicx}
\usepackage{dcolumn}
\usepackage{bm}
\usepackage{longtable}
\usepackage{amsfonts,amsmath,amssymb}
\usepackage{color}

\usepackage{ulem}

\begin{document}

\title{First-principles study of hydrogen-bonded molecular conductor $\kappa$-H$_3$(Cat-EDT-TTF/ST)$_2$}  

\author{Takao Tsumuraya$^{1,2}$\thanks{Email address: TSUMURAYA.Takao@nims.go.jp}
Hitoshi Seo$^{3, 4}$,
Reizo  Kato$^{1}$, 
and Tsuyoshi Miyazaki$^{2}$}
\affiliation{
$^{1}$Condensed Molecular Materials Laboratory, RIKEN, Wako-shi, Saitama  351-0198, Japan\\
$^{2}$Computational Materials Science Unit, National Institute for Materials Science, Tsukuba 305-0044, Japan\\
$^{3}$Condensed Matter Theory Laboratory, RIKEN, Wako-shi, Saitama 351-0198, Japan\\
$^{4}$Quantum Matter Theory Research Team, CEMS, RIKEN, Wako-shi, Saitama 351-0198, Japan\\
}
\date{\today}
\begin{abstract}
We theoretically study hydrogen-bonded molecular conductors synthesized recently, 
$\kappa$-H$_3$(Cat-EDT-TTF)$_2$ and its diselena analog, 
$\kappa$-H$_3$(Cat-EDT-ST)$_2$, by first-principles density-functional theory calculations.
In these crystals, two H(Cat-EDT-TTF/ST) units share a hydrogen atom with a short O--H--O hydrogen bond. 
The calculated band structure near the Fermi level shows a quasi-two-dimensional character, 
with a rather large interlayer dispersion due to the absence of insulating layers in contrast with
conventional molecular conductors. We discuss effective low-energy models based on H(Cat-EDT-TTF/ST) units and its dimers, respectively, where the microscopic character of the orbitals composing them are analyzed. 
Furthermore, we find a stable structure which is different from the experimentally determined structure, where the shared hydrogen atom becomes localized to one of the oxygen atoms, in which charge disproportionation between the two types of H(Cat-EDT-TTF) units is associated. 
The calculated potential energy surface for the H atom is very shallow near the minimum points, 
therefore the probability of the H atom can be delocalized between the two O atoms. 
\end{abstract}

\pacs{Valid PACS appear here}
\keywords{First-principles calculation, hydrogen bonds}
\maketitle
\section{INTRODUCTION}
Molecular conductors---molecular solids 
which possess electron/hole carriers---exhibit a rich variety of physical properties, such as 
superconductivity, Mott transition, charge ordering, magnetic ordering, 
quantum spin liquid (QSL) state, 
and so on.~\cite{OrganicSC_book08, Kanoda_Kato_2011ARCMP} 
In most of them, the compounds consist of 
 two~(or more) kinds of molecular species so that the molecules 
 can have partial valence states 
 which give rise to partially occupied electronic bands at the Fermi level. 
The emergent phenomena can be associated with the geometries of molecular packing 
 which determine the electronic structure, combined 
 with the strong electron-electron and electron-phonon interactions.
In order to study them theoretically,
 the tight-binding model, constructed by taking \textit{molecular orbitals} as basis functions,~\cite{T_Mori_Bull_ChemSocJpn84} 
 is shown to be valid as a starting point to describe the band structure near the Fermi level;
 then the interaction effects can be taken into account by Hubbard-type models.~\cite{KinoFukuyama_dimer96} 
Such model calculations have successfully been applied to various families 
and offered not only a description of each phenomenon, but also a systematical view 
 between the materials.~\cite{SeoHott_ChmRev04}
In recent years, 
 the tight-binding parameters in such models, 
 i.e., the intermolecular transfer integrals,  
 can qualitatively be evaluated by first-principles density functional theory~(DFT) calculations.
 This approach is efficient, especially for newly developed molecular systems where semiempirical approaches are not always assured.~\cite{Miyazaki_PRB96_DCNQI, Seo_JPSJ08_frag, Kino_Beta_ET2Cl2, k_ET_KNakamura_JPSJ09, Seo_Pd_dmit15}

\begin{figure}[t, b]
\begin{center}
\includegraphics[width=1.0\linewidth]{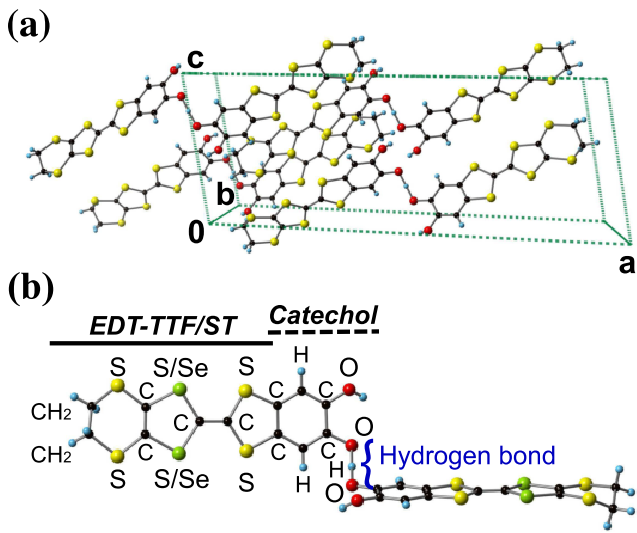}
\end{center}
\caption{(Color online) (a) Crystal structure of $\kappa$-type H$_3$(Cat-EDT-TTF)$_2$ ($\kappa$-S) measured at 50~K. (b) Molecular structure of H$_3$(Cat-EDT-TTF/ST)$_2$. 
}
\setlength\abovecaptionskip{0pt}
\label{Crystal}
\end{figure}

\begin{figure}[t]
\begin{center}
\includegraphics[width=1.0\linewidth]{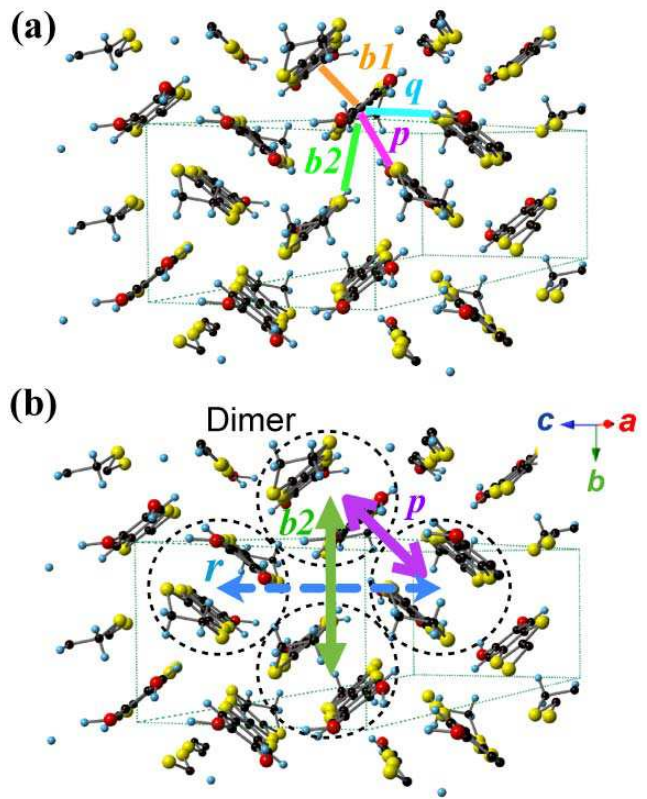}
\end{center}
\caption{(Color online) View of the crystal structure along the $a$ axis showing the $bc$ plane. Definitions of (a) inter-molecular and (b) interdimer bonds are presented.} \setlength\abovecaptionskip{0pt}
\label{Crystal2}
\end{figure}

Recently, a new class of molecular conductors containing 
hydrogen bonds has been developed. 
The compounds are based on catechol with ethylenedithiote-tetrathiafulvalene 
molecules, Cat-EDT-TTF, and its diselena analog ethylenedithiote-selenathiafulvalene, 
Cat-EDT-ST.~\cite{Kamo12_H3_Cat}    
Among them, $\kappa$-H$_3$(Cat-EDT-TTF)$_2$ (space group: $C$2/$c$, abbreviated here as $\kappa$-S) 
and its isostructural analog $\kappa$-H$_3$(Cat-EDT-ST)$_2$ ($\kappa$-Se) are attracting interest, whose crystal structure is shown in Fig.~\ref{Crystal}(a).~\cite{Isono_H3Cat_2013} 
Their structural properties are unique where the H(Cat-EDT-TTF/ST) units are stacked two~dimensionally (2D) in the $bc$~plane, connected by the hydrogen atom in the interlayer $a$~direction. 
In the 2D plane, the H(Cat-EDT-TTF/ST) units are arranged in the so-called $\kappa$-type packing, 
where face-to-face dimers are arranged in an anisotropic lattice~(Fig.~\ref{Crystal2}). 

$\kappa$-S shows a Mott insulating behavior where localized $S$ = 1/2  spins appear and the possible realization of a QSL state was indicated. 
Despite the exchange interaction of about 80 -- 100 K, 
the magnetic susceptibility shows no magnetic ordering down to 2~K
and the magnetic torque shows a paramagnetic behavior even at 50~mK.~\cite{Isono_QSL_H3Cat13}  
It is also reported that a deuterium~(D) substitution for $\kappa$-S 
induces a non--magnetic charge disproportionated state 
associated with D localization.~\cite{A_Ueda_DCat}
On the other hand, 
$\kappa$-Se is also an insulator, but shows metallic behavior at room temperature 
by the application of pressure above 1.3 GPa, 
where the room-temperature electric conductivity increases up to 180~S/cm at 2.2 GPa; 
$\kappa$-S remains insulating up to about 1.6~GPa. 
This conductivity observed in $\kappa$-Se is the highest among molecular conductors.~\cite{Isono_H3Cat_2013} 

In previous reports,~\cite{Isono_H3Cat_2013, Isono_QSL_H3Cat13} 
their electronic properties were discussed by an analogy to conventional molecular conductors such as~$\kappa$-(BEDT-TTF)$_2X$.~\cite{Miyagawa_ChemRev}
The dimers of H(Cat-EDT-TTF/ST) units form the same 2D structure as $\kappa$-(BEDT-TTF)$_2$$X$. 
The expected oxidized state of each unit is +0.5, which is also the case for BEDT-TTF$^{+0.5}$. 
There, the dimer formation and the effect of strong electron correlation are known to make the system Mott insulating (so-called dimer-Mott insulator), resulting in each dimer carrying a localized spin $S$ = 1/2. 
Then, the QSL state behavior in $\kappa$-S can be discussed based on the analogy to 
other molecular QSL candidates, including $\kappa$-(BEDT-TTF)$_2$Cu$_2$(CN)$_3$.~\cite{Kanoda_Kato_2011ARCMP} 

However, differently from such conventional molecular conductors, 
these title compounds do not have discrete insulating layers, 
which play a role as the charge reservoir for the electrical conduction in conventional systems. 
Instead, the constituent molecule H$_3$(Cat-EDT-TTF/ST)$_2$ has a hydrogen atom shared by two H(Cat-EDT-TTF/ST) units, as shown in Fig.~\ref{Crystal}(b). 
In Refs.~\onlinecite{Isono_H3Cat_2013} and~\onlinecite{Isono_QSL_H3Cat13}, their electronic structures are discussed by assuming the usual tight-binding model, whereas the interlayer direction is neglected. 
However, the character of the orbitals which constitute the low-energy effective model, i.e., the appropriate unit whose orbital can describe the character of bands near the Fermi level, remains unclear.
Therefore it is important to study these materials from first-principles calculations and derive effective models. 

Another interest here is the role of the hydrogen bonds. 
Two H(Cat-EDT-TTF/ST) units share a hydrogen atom 
with a relatively short O--H--O hydrogen bond~(e.g.,~O$\cdots$O~lengths of $\kappa$-S and $\kappa$-Se are 2.45~and 2.44~\AA, respectively). 
The shared H atom is at the center of two nearest O atoms, with O--H lengths of 1.23~and~1.22~\AA~for $\kappa$-S and $\kappa$-Se, respectively.~\cite{Isono_H3Cat_2013}
These geometries are in accordance with the systematic survey of the correlation between O--H and 
O$\cdots$O lengths in hydrogen bonds for a large number of material systems.~\cite{H_posi08, OH_OO_Ichi78, OOSteiner94}  
Proton dynamics in such short intermolecular hydrogen bonds play essential roles 
in various functional molecular materials such as organic ferroelectrics~\cite{F_Ishii06_organicFE, PhzH2ca12, Horiuchi_H-bond11, LeeVDB12}, 
molecular conductors~\cite{Mitani_HB88, Akutagawa02, Zwitterionic2010}, and biochemical reactions.~\cite{HtransPhotochro94, Tapia_enzym94} 
In general, the shape of the potential for hydrogen in short hydrogen bonds 
is sensitive to the O$\cdots$O distance.~\cite{Cleland_Scie94_LBHB, LBHB_Science94, Dmarx_CPC05, HamadaMorikawaPRB10}  
For such systems, the position and its influence to the relevant electronic structure are crucially important.
However, it is difficult to determine the position of the H atom using neutron scattering techniques 
for the title compounds due to their small sample size. 
First-principles calculation is a useful tool to clarify such properties. 

In this work, we investigate the electronic and structural properties of $\kappa$-S and $\kappa$-Se 
by first-principles DFT calculations. 
We report unique properties of the electronic structure compared with conventional molecular conductors, 
as well as the differences between these two compounds. 
We evaluate transfer integrals by fitting the DFT band structure 
and discuss the character of orbitals composing the model by analyzing the bands near the Fermi energy. 
Regarding the hydrogen bonds, we calculate the potential surface to 
investigate the degree of localization of the shared H atom, and discuss the stability of the structures. 

\section{COMPUTATIONAL DETAILS}
The first-principles DFT calculations were performed by an all-electron full-potential linearized augmented plane wave (FLAPW) method as implemented in the QMD-FLAPW12 code.~\cite{Wimmer1981, L_KA, Weinert} 
We used the exchange-correlation functional of the generalized gradient approximation~(GGA) by Perdew-Burke-Ernzerhof;~\cite{GGA_PBE}
non-spin-polarized calculations were performed. 
We used plane-wave cutoff of $|\bf{k} + \bf{G}|$ $\leq$ 5.5 a.u.$^{-1}$ ($\sim$30 Ry) and 14.5 a.u.$^{-1}$~($\sim$210 Ry) for LAPW basis functions, and charge density and potential, respectively.
For a given $\bf{k}$~point, the LAPW basis set includes all reciprocal-lattice vector $\bf{G}$ within the cutoff energy. 
Uniform $\bf{k}$-point meshes of 5$\times$5$\times$3 were used for Brillouin zone integrations.~\cite{prim} 
Common muffin-tin (MT) sphere radii are set to be 1.98 for S,  2.20 for Se, 1.18 for O, 1.28 for C, and 0.7 a.u. for H atoms. The lattice harmonics with angular momenta up to $l$ = 8 are set for all the atoms. 

The title compounds are composed of light-elements with relatively short bond lengths, and thus their MT spheres are set to smaller values than inorganic materials such as oxides. Although this situation requires a careful consideration also for the numerical parameters such as plane-wave cutoffs and $\bf{k}$~points, anisotropic molecular crystals can be treated precisely with the strict parameters used in the present study. 
Due to the progress of computational resources and developments of parallelization techniques, the accurate all-electron FLAPW method is now available for molecular solids.~\cite{Min_Freeman_H2, PhysRevLett.103.067004, K_Nakamura_Fe_phthalocyanine, P3HT_Tsumu, Ferber_k_ET14}

We used experimental lattice parameters of $\kappa$-S and $\kappa$-Se 
measured at 50~K and 30~K, respectively, throughout this paper. ~\cite{Isono_H3Cat_2013} 
The lattice parameters are listed in Table~\ref{Lattice parameters}. 
For the optimization of the internal coordinates, we take several different approaches depending on the purpose as follows:
 
In Sec. III A,~we used experimental atomic positions except for the unshared H atoms. 
The positions of the H atoms were optimized theoretically because the experimental positions determined by x-ray diffraction measurements give unphysically short O--H and C--H distances with the chemical bonds. 

In Sec.~III B,~we optimized all the atomic positions starting from the experimental positions. 
We also performed the optimization with the shared H atom being slightly displaced ($\sim$0.1~\AA) from the center of O$\cdots$O and confirmed that the optimized position is recovered, showing the $C$2/$c$ symmetry. 
Furthermore, we took the initial position of the shared hydrogen atom at a position more closely to one of the two nearest O atoms (the displacement from the center:~$\sim$0.24~\AA), then all the atomic positions were optimized. As a result, we found another structure with $P\bar{1}$ symmetry. 
In all the calculations, the atomic positions were relaxed until the residual maximum force is smaller than 10$^{-3}$ hartree/Bohr. 

In Sec. III~C,~we calculate the change of the total energy as a function of the position of the shared H atom for the two optimized structures described in Sec.~III B.
In this calculation, the atomic positions except for the shared H atom were fixed, 
because we consider the case where other atoms cannot respond to the movement of the H atom.

In Sec. III~A, both $\kappa$-S and $\kappa$-Se are discussed, while in Secs. III~B and III~C, only $\kappa$-S is treated.

\begin{table}[b]
\caption{Lattice parameters of $\kappa$-H$_3$(Cat-EDT-TTF/ST)$_2$ with the space group $C$2/$c$.\cite{Isono_H3Cat_2013}}
\label{Lattice parameters}
\begin{center}
\begin{tabular}{lccccccc}
\hline
 & $a$~(\AA) & $b$~(\AA) & $c$~(\AA) & $\beta$~[deg.]  \\
\hline
$\kappa$-S &29.43 & 8.36 & 11.13 & 100.92  \\ 
$\kappa$-Se & 29.88 & 8.37 & 11.12 & 101.11 \\ 
\hline
\end{tabular}
\end{center}
\end{table}

\begin{figure}[t]
\begin{center}
\includegraphics[width=1.0\linewidth]{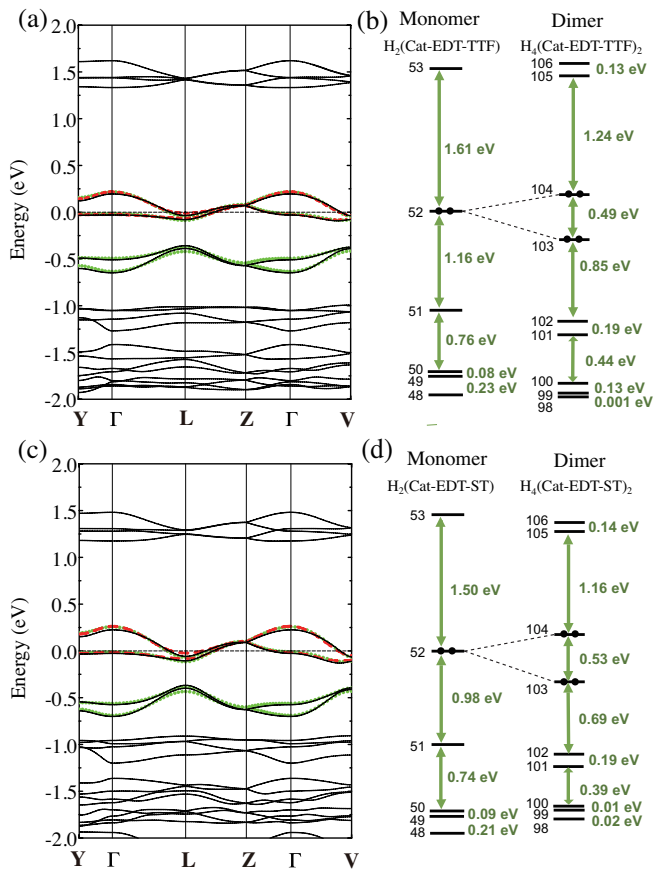}
\end{center}
\caption{(Color online) (a) First-principles band structure of $\kappa$-H$_3$(Cat-EDT-TTF)$_2$. (b) Energy diagram of isolated monomer H$_2$(Cat-EDT-TTF) and its dimer, H$_4$(Cat-EDT-TTF)$_2$, where all the O--H bonds were set to be 1.0 \AA.
(c) First-principles band structure of $\kappa$-H$_3$(Cat-EDT-ST)$_2$. (d) Schematic energy diagram of isolated monomer H$_2$(Cat-EDT-ST) and its dimer, H$_4$(Cat-EDT-ST)$_2$. 
In both band structures (a) and (c), dashed curves show the fitted results for the four-band model, and dotted curves show the fitted results of the two-band (dimer) model. 
The origin of the vertical axis with a dashed line shows the Fermi level.}
\setlength\abovecaptionskip{0pt}
\label{band_H3Cat}
\end{figure}

\section{RESULTS AND DISCUSSIONS}
{\subsection{Electronic structure of $\kappa$-S and $\kappa$-Se using experimental positions with $C2$/$c$ symmetry}
Figures~\ref{band_H3Cat}(a) and (c) show the calculated band structures 
of $\kappa$-S and $\kappa$-Se, respectively, where both show quasi-2D electronic structures. 
The calculated bands show a four-band structure in the energy range of $-$0.7 to +0.25 eV. 
There are two bands each separated by an energy gap.
The upper two bands cross the Fermi energy; they are half-filled.
These features are indeed analogous to 
$\kappa$-(BEDT-TTF)$_2$$X$ as mentioned in Sec. I.

Comparing the band structures between $\kappa$-S and $\kappa$-Se in Fig.~\ref{band_H3Cat}, 
the widths of the two bands crossing the Fermi level are markedly different,~i.e.,~$\kappa$-Se has a larger bandwidth of 359 meV than that of $\kappa$-S, 312 meV. This comes from the delocalized character of Se--$p$ orbitals than S--$p$ orbitals. 
A notable point of their electronic structure is that 
the bands along the $Y$--$\Gamma$ direction which correspond to the interlayer direction ($a$~axis) are rather dispersive, indicating the existence of a relatively large three~dimensionality. 
This originates from the absence of insulating layers between electron conducting layers, 
peculiar to these new materials, which is also seen in the estimated transfer integrals as in the following. 

Now let us consider effective tight-binding models. 
From the analogy to $\kappa$-(BEDT-TTF)$_2$$X$, 
we assume each basis function to be located on the monomers of the H(Cat-EDT-TTF/ST) unit and its dimer, respectively. 
Their definitions for the transfer integrals are illustrated in Figs.~\ref{Crystal2}(a) and (b), respectively.  
The four-band model is the same model as discussed in Ref.~\onlinecite{Isono_H3Cat_2013} where the H(Cat-EDT-TTF/ST) units are taken as the lattice points, 
while we also consider a half-filled two-band model taking the face-to-face dimer of H(Cat-EDT-TTF/ST) units as lattice sites, as frequently discussed in molecular conductors with dimerization.~\cite{KinoFukuyama_dimer96,SeoHott_ChmRev04} 
In fact, the calculated eigenfunctions for isolated units shown in Figs.~\ref{band_H3Cat}(b) and (d) are consistent with such models. The highest occupied molecular orbital~(HOMO) of a monomer locates near the center of the four bands, while its dimer shows two levels located at the upper and lower bands. The former (latter) consists of an antibonding (bonding) combination of the monomer wavefunctions. 
Note that the isolated units for the calculation here of the molecular orbitals are neutral H$_2$(Cat-EDT-TTF/ST) and its dimer. We will discuss this point later.}

\begin{figure}[t,b]
\begin{center}
\includegraphics[width=1.0\linewidth]{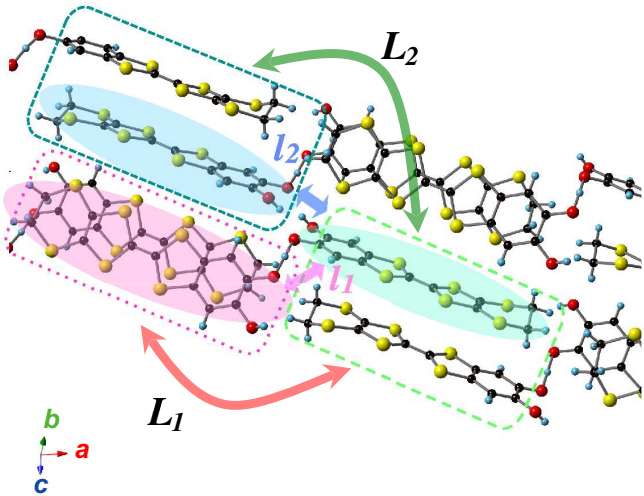}
\end{center}
\caption{(Color online) Notation of interlayer bonds of $\kappa$-H$_3$(Cat-EDT-TTF)$_2$. 
$l_1$ and $l_2$ are those between the H(Cat-EDT-TTF) units, whose former is that connected by the shared H atom. 
$L_1$ and $L_2$ denote the two different interdimer bonds.}
\setlength\abovecaptionskip{-5pt}
\label{L1L2}
\end{figure}

In Table~\ref{Trans} we list the transfer integrals obtained by the tight-binding fittings. 
For $\kappa$-S, the transfer integrals obtained by the four-bands fitting to the DFT bands agree well with those obtained by the extended H\"{u}ckel method (the interlayer bonds were not considered in Ref~\onlinecite{Isono_H3Cat_2013}). 
On the other hand, for $\kappa$-Se, there are quantitative differences between them. 
In both compounds, a transfer integral $b1$, which corresponds to the face-to-face overlaps of the units within the dimers, is the largest. 
This justifies the approximate mapping to the two-band model, where the basis functions are taken as an antibonding combination of the basis sets.~\cite{KinoFukuyama_dimer96,SeoHott_ChmRev04}
The second largest $b2$, where the overlap of C--$p_z$~and~S--$p_z$ orbitals in one unit with those in the nearest dimer exists, mainly determines the bandwidth. 
The three-dimensionality mentioned above is seen in the interlayer transfer integrals; they are not negligible compared with intralayer values. 
Comparable values of $\sim$10 -- 20 meV are found along two interlayer bonds $l1$ and $l2$ (see Fig.~\ref{L1L2}) in both compounds. 
These two are comparable to an intralayer $q$, and were needed for a reasonable fitting to the dispersive bands along the $Y$--$\Gamma$ direction. 
In $\kappa$-Se, the transfer integral along $b2$ is larger than that in $\kappa$-S; this is the main reason for the larger bandwidth of $\kappa$-Se than $\kappa$-S. 

Among the transfer integrals obtained by the two-band fitting, the largest two, those along $b2$ and $p$, form an anisotropic triangular lattice, similar to the case of $\kappa$-(BEDT-TTF)$_2$$X$.~\cite{KinoFukuyama_dimer96, SeoHott_ChmRev04}   
We find that the value along one side~($b2$) of an isosceles triangle is larger than that along the two equal sides~($p$). 
The ratio of our interdimer transfer integrals, ${b2}$/$p$ which is known as $t^{\prime}$/$t$ in the literature applied to different compounds,~\cite{Kanoda_Kato_2011ARCMP} is estimated as 1.25 and 1.42 for $\kappa$-S and $\kappa$-Se, respectively. 
The extended H\"{u}ckel calculations give the ratio of 1.48 and 1.66, respectively.~\cite{Isono_H3Cat_2013} \
Then, one view of the electronic structure is that one-dimensional chains formed by the $b2$ bonds are connected by frustrated interchain couplings along the $p$ bonds. 
A similar tendency was reported in first-principles studies for $\beta^{\prime}$-EtMe$_3$Sb[Pd(dmit)$_2$]$_2$ which also shows a QSL.~\cite{Tsumu_Pd_dmit2_13, Nakamura, Jacko_dmitPRB14}  
The relatively large interlayer transfer integrals are also obtained by the two-bands fitting showing two comparable absolute values of $\sim$10~meV along $L_1$ and $L_2$ bonds~(see Fig.~\ref{L1L2}), which is expected to act as an interlayer spin-frustration effect. 
\begin{table}
\caption{Intermolecular~(four-band model) and interdimer~(two-band model) transfer integrals obtained by fitting to first-principles band structure. The unit is meV.
The notation of intermolecular/interdimer bonds within the 2D plane is shown in Fig.~\ref{Crystal2}, and that for interlayer directions is shown in Fig.~\ref{L1L2}. 
}
\label{Trans}
\begin{center}
\begin{tabular}{lccccccccc}
\hline\hline
Intermolecule & & $b1$ & $b2$ & $p$ & $q$ & $l1$ & $l2$\\
\hline
$\kappa$-S & DFT & 241 & 75 & 40 & 11 & 14 & 17 \\
& H\"{u}ckel~\cite{Isono_H3Cat_2013} & 230 & 85 & 46 & 12 \\ 
\hline
$\kappa$-Se & DFT & 258 & 98 & 42 & 18 & 14 & 19 \\
& H\"{u}ckel~\cite{Isono_H3Cat_2013} & 318 & 140 & 63 & 21 \\ 
\hline
\\
\hline
Interdimer &  & $b2$ & $p$ & $r$ & ${b2}$/$p$ & $L1$ &  $L2$\\
\hline
$\kappa$-S & DFT & 32 & -26 & -4 & 1.25 & -9 & 10 \\
\hline
$\kappa$-Se & DFT & 44 & -31 & -5 & 1.42 & -9 & 11 \\
\hline\hline
\end{tabular}
\end{center}
\end{table}

\begin{figure*}[t]
\begin{center}
\includegraphics[width=1.0\linewidth]{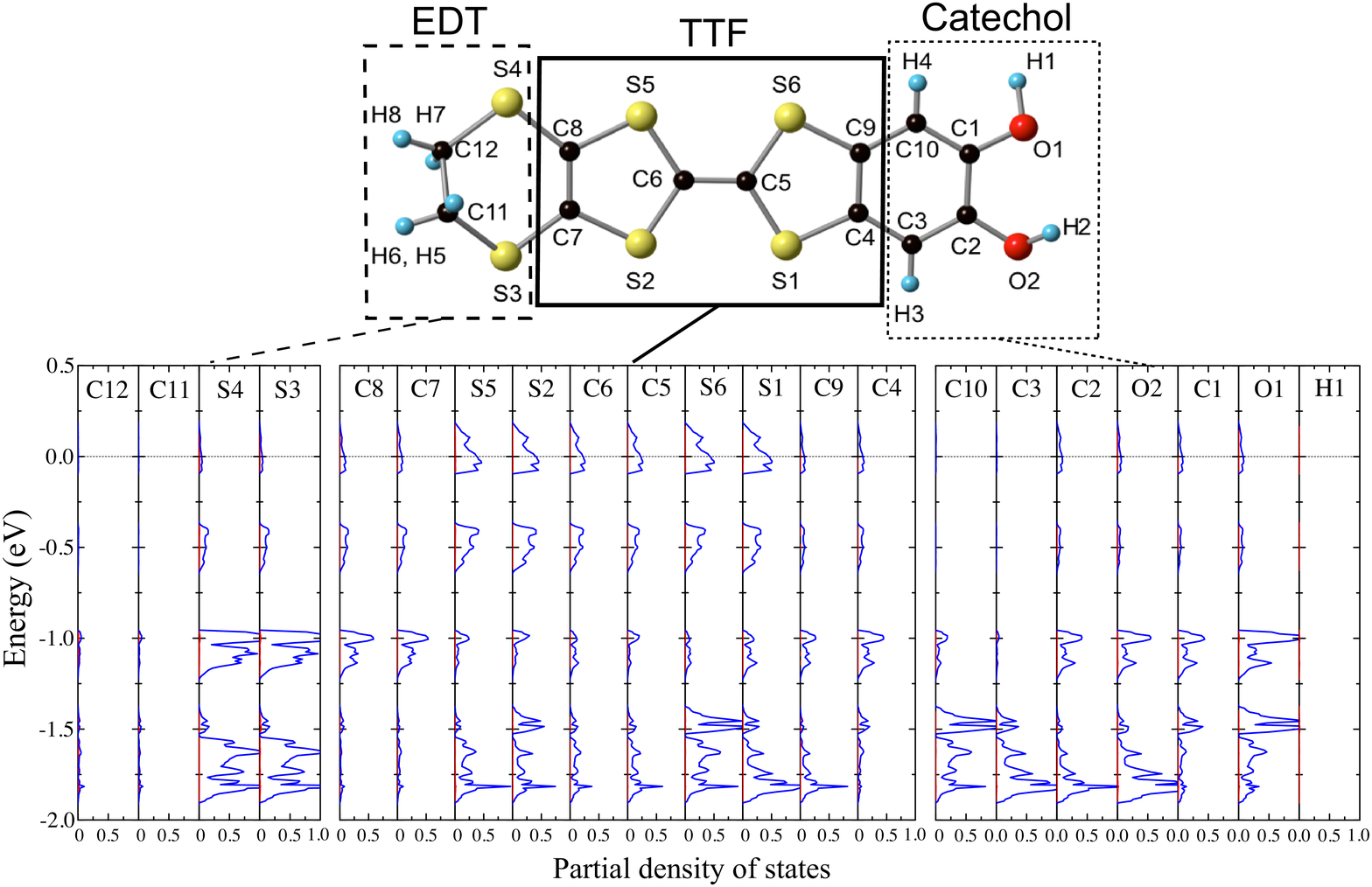}
\end{center}
\caption{(Color online) Partial density of states of $\kappa$-H$_3$(Cat-EDT-TTF)$_2$ with the space group $C$2/$c$. The atom assignments are shown in the figure (see also Table~\ref{C2c_opt}).  
The origin of the vertical axis with a dashed line shows the Fermi level.}
\setlength\abovecaptionskip{-5pt}
\label{pdos_H3Cat}
\end{figure*}

Next, we discuss the character of Kohn-Sham orbitals near the Fermi level, in order to obtain atomic scale information about the low energy electronic structure. 
Figure~\ref{pdos_H3Cat} shows the partial density of states~(DOS) 
projected on each orbital in $\kappa$-S within the MT sphere. 
The bands crossing the Fermi level are predominantly composed of the $p$~states of C and S atoms belonging to the TTF~part as plotted in Fig.~\ref{pdos_H3Cat}.  
On the other hand, the wavefunctions originating from C and S atoms 
of ethylenedithiote~(EDT) and Catechol parts appear at the lower energy 
region between --1.25 and --0.9~eV. 
These results indicate that the low-energy properties are mainly governed by the TTF part, 
which constitutes the basis functions for the effective model discussed above.

In previous reports,~\cite{Isono_H3Cat_2013, Isono_QSL_H3Cat13} the eigenstates near the Fermi level were discussed based on the singly occupied molecular orbital (SOMO) of an isolated H$_3$(Cat-EDT-TTF)$_2$ molecule shown in Fig.~\ref{Crystal}(b). 
However, since the SOMO is delocalized over the entire Cat-EDT-TTF molecule by the $\pi$~conjunction, the character of the eigenstates near the Fermi level for the solid [Figs.~\ref{band_H3Cat}(a) and~\ref{band_H3Cat}(c)] is very different from that of this SOMO. 
In fact, the eigen energies of H$_3$(Cat-EDT-TTF)$_2$ molecule in Fig.~\ref{Crystal}(b) do not correspond to the energy bands in Fig.~\ref{band_H3Cat}(a). 
On the other hand, as shown in Figs.~\ref{band_H3Cat}(b) and~\ref{band_H3Cat}(c), the eigenstates of the H$_2$(Cat-EDT-TTF) unit and its face-to-face dimer that are described above, showing good agreement with the band structure, indeed possess a large contribution from the TTF part. 

\subsection{Theoretical relaxations of atomic positions with lowered symmetry}
\begin{figure}
\begin{center}
\includegraphics[width=1.0\linewidth]{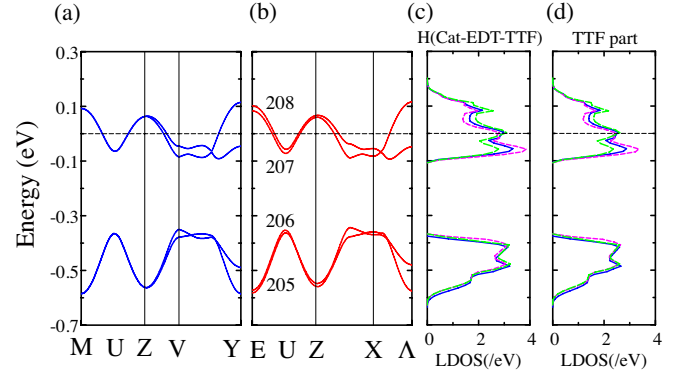}
\end{center}
\caption{(Color online)~Band structures of optimized (a) $C$2/$c$ and (b) H-localized ($P\bar{1}$) structures of $\kappa$-H$_3$(Cat-EDT-TTF)$_2$.  
To see how the degenerated bands split by the symmetry lowering, the band structure along the common symmetric line is plotted [different from Fig.~\ref{band_H3Cat}(a)]. 
(c) Local density of states (LDOS) of the H(Cat-EDT-TTF) with $C$2/$c$~(solid curve) and $P\bar{1}$ structures. The later contains two parts, H$_2$(Cat-EDT-TTF)~(dashed curve) and H(Cat-EDT-TTF)~(broken curve). 
(d) LDOS of the TTF part, which is summation of the partial DOS from C--$p$, S--$p$, O--$p$, and H--$s$ states belonging to each TTF part. 
The origin of the vertical axis with a dashed line shows the Fermi level.}
\setlength\abovecaptionskip{0pt}
\label{bands_P1bar}
\end{figure}

Next let us turn to the role of the hydrogen bonds, which is characteristic 
in these molecular conductors. 
The position of the H atom determined by x-ray diffraction measurements at the center of O$\cdots$O may be the averaged one due to their thermal or quantum fluctuations.
In this section, we study the possibility that the H atom binds to one of the two O atoms in spite of the experimental results. 
We first optimized the atomic positions of all atoms with the experimental symmetry $C$2/$c$, and their fractional coordinates are listed in Table~\ref{C2c_opt}.
Note that all the H(Cat-EDT-TTF) units sharing a H atom are crystallographically equivalent with this symmetry.
The calculated structural parameters are shown in Table~\ref{OHgeometry}.
Although there are small differences between the experimental and optimized structures with the $C$2/$c$ symmetry, we have confirmed that the calculated band structure using the optimized structure, shown in Fig.~\ref{bands_P1bar}(a) is almost the same as the band structure for the experimental structure [Fig.~\ref{band_H3Cat}(a)], implying that the optimized structure with the $C$2/$c$ symmetry is very close to the experimental structure.~[The symmetric lines in Figs.~\ref{band_H3Cat}(a) and ~\ref{bands_P1bar}(a) are different.]
\begin{table}
\caption{Theoretically optimized atomic positions for the $C$2/$c$ structure of $\kappa$-H$_3$(Cat-EDT-TTF)$_2$.
 All the H(Cat- EDT-TTF) units in the unit cell are crystallographically equivalent.}
\label{C2c_opt}
\begin{center}
\begin{tabular}{lcccc}
\hline\hline
Label & \multicolumn{3}{c}{Fractional coordinates} \\  
 & $x$ & $y$ & $z$ \\
\hline
O1 & 0.0131 & 0.0008 & 0.1528 \\
O2 & 0.4600 & 0.7035 & 0.0179 \\
H1 & 0.0000 & 0.0095 & 0.2500 \\
H2 & 0.9838 & 0.1203 & 0.0511 \\
H3 & 0.1260 & 0.7153 & 0.0262 \\
H4 & 0.0809 & 0.9057 & 0.8295 \\
H5 & 0.4453 & 0.9727 & 0.0168 \\
H6 & 0.0186 & 0.3645 & 0.9203 \\
H7 & 0.0643 & 0.6617 & 0.2513 \\
H8 & 0.4599 & 0.9655 & 0.2461 \\
C1 & 0.0575 & 0.9572 & 0.1663 \\
C2 & 0.5712 & 0.3520 & 0.0786 \\
C3 & 0.1167 & 0.7977 & 0.0934 \\
C4 & 0.1494 & 0.8497 & 0.1932 \\
C5 & 0.2238 & 0.9022 & 0.3520  \\
C6 & 0.2684 & 0.1014 & 0.9169 \\
C7 & 0.6441 & 0.8464 & 0.0125 \\
C8 & 0.3447 & 0.0448 & 0.0681 \\
C9 & 0.1367 & 0.9579 & 0.2772 \\
C10 & 0.0911 & 0.0107 & 0.2643 \\
C11 & 0.0534 & 0.4169 & 0.9297 \\
C12 & 0.4351 & 0.0512 & 0.1953 \\
S1 & 0.2070 & 0.7868 & 0.2205 \\
S2 & 0.6893 & 0.7781 & 0.1270 \\
S3 & 0.0897 & 0.2658 & 0.0168 \\
S4 & 0.3800 & 0.9508 & 0.1919 \\
S5 & 0.2868 & 0.9812 & 0.0451 \\
S6 & 0.1810 & 0.9795 & 0.8959 \\
\hline\hline
\end{tabular}
\end{center}
\end{table}
\begin{table*}
\caption{Theoretically optimized and experimentally determined structural parameters around hydrogen bonds in $\kappa$-H$_3$(Cat-EDT-TTF)$_2$. 
The atom indexes are defined in Fig.~\ref{Crystal_Hbonds}. In the $C$2/$c$ structures, O1 and O3 atoms, C1 and C3 atoms are crystallographically equivalent.}
\label{OHgeometry}
\begin{center}
\begin{tabular}{lcccccccc}
\hline
\hline
& Space group & $d$(O$\cdots$O) & $d$(O--H) & $\angle$O--H--O& $d$(O1--C1) & $d$(O3--C3) & $d$(O1--O2) & $d$(O3--O4)\\
&  & [\AA] & [\AA] & [deg.] & [\AA] &  [\AA] & [\AA] & [\AA]\\
\hline
DFT calc. & $C$2/$c$ & 2.43 & 1.22 & 173.2 & 1.34 & 1.34 & 2.78 & 2.78\\
DFT calc. & $P\bar{1}$ & 2.44 & 1.13/1.31 & 173.0 & 1.35 & 1.32 & 2.76 & 2.79\\
Experiments~\cite{Isono_H3Cat_2013}  & $C$2/$c$ & 2.45 & 1.23 & 168.3 & 1.35 & 1.35 & 2.77 & 2.77\\
\hline
\hline
\end{tabular}
\end{center}
*Experimental atomic positions were determined by x-ray diffraction method at 50~K without any DFT optimizations.~\cite{Isono_H3Cat_2013} 
\end{table*}

In order to investigate whether the central position of
the shared H atom is stabilized just because of the constraint in the $C$2/$c$ symmetry,
we varied the initial condition of the optimization where the H atom is displaced from the center and performed 
the structure optimization with a lower symmetry $P\bar{1}$ by removing
the glide symmetry of $C$2/$c$. 
Here, we assumed two sizes of displacement of the H atom as initial conditions of the geometric optimization. 
First, we displaced the H atom by 0.1~\AA,~and then relaxed all of the atomic positions. 
The optimized structure is found to be the same as the optimized structure with $C$2/$c$ symmetry. 
This means that the $C$2/$c$ structure is at least one of the local-minimum structures even within the classical treatment of an H atom. 

\begin{figure}
\begin{center}
\includegraphics[width=1.0\linewidth]{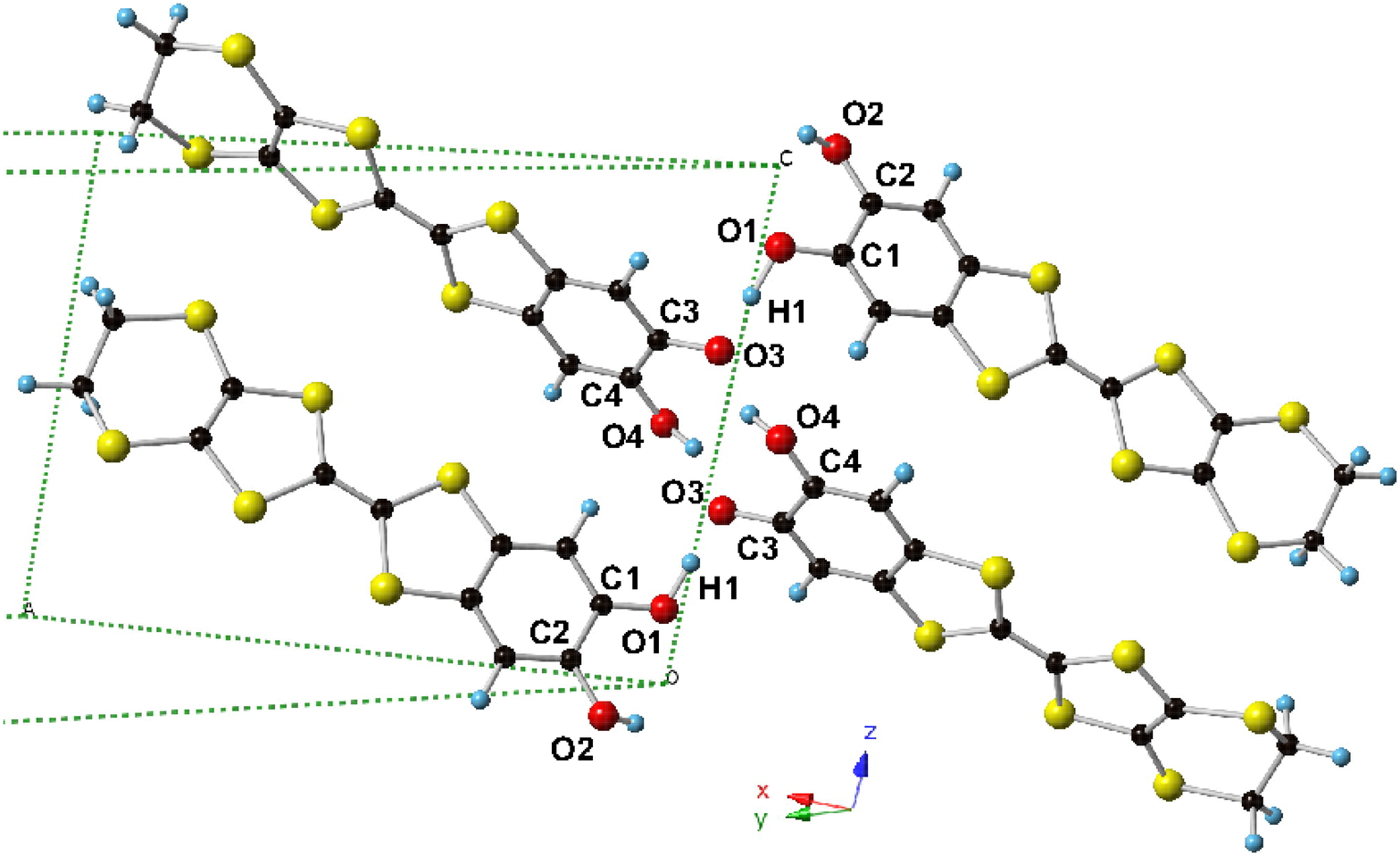}
\end{center}
\caption{(Color online) Structural properties around hydrogen bonds in H-localized structure~($P\bar{1}$) of $\kappa$- H$_3$(Cat-EDT-TTF)$_2$. 
The shared H atom (H1) is located at the fractional coordinate of (0.010, --0.008, 0.241)~[(0.000, --0.010, 0.250) in the optimized $C$2/$c$ structure].}
\setlength\abovecaptionskip{0pt}
\label{Crystal_Hbonds}
\end{figure}

Next, to explore the possibility that the H atom is localized, we considered an initial position of the H atom which is closer to O1~(the O1--H1 distance was set to be 1.0~{\AA}), and then relaxed the positions of all the atoms again. 
As a result, we found another local minimum structure where H is localized near O1 (Fig.~\ref{Crystal_Hbonds} and Table~\ref{P1bar_opt}).
As shown in Table~\ref{OHgeometry}, this structure has two inequivalent O-H distances,~1.13~and~1.31~\AA,
and now there exist two distinct molecules H$_2$(Cat-EDT-TTF) and H(Cat-EDT-TTF),
abbreviated here as w-H and w/o-H units, respectively. 
We found that this structure is energetically comparable to the optimized $C$2/$c$ structure (lower by 2 meV), 
which we refer to as the $P\bar{1}$ structure in the following. 

Let us compare the electronic properties for the two different structures. 
The band structures of the optimized $C$2/$c$ and $P\bar{1}$ structures are shown in Fig.~\ref{bands_P1bar}(a) and (b), respectively, 
where the symmetric lines are those for the $P\bar{1}$ structure for comparison. 
One can see the splitting of the bands near the Fermi level along the M~(E)--U--Z line when the symmetry is lowered from $C$2/$c$ to $P\bar{1}$. 
By analyzing the Kohn-Sham wavefunctions~(at the M~(E), U and Z-points), the bands with index 205 and 208 in Fig~\ref{bands_P1bar}(b) are mainly composed of bonding and antibonding states of the dimers of w/o-H units and those with 206 and 207 are those of dimers of the w-H units. 

The comparison of the local density of states~(LDOS) also indicates a difference of the electronic structure~[Fig.~\ref{bands_P1bar}(c)].  
In the occupied state of the energy range around --0.1 --~0 eV, compared to the case of the $C$2/$c$ structure where all the H(Cat-EDT-TTF) units are equivalent, the LDOS for the w-H unit increases but that of the w/o-H unit decreases. 
On the other hand, above the Fermi level from 0 to +~0.2~eV, the LDOS shows the opposite behavior. 
This result indicates the existence of a small amount of charge transfer from the w/o-H to the w-H unit. 
The LDOS which is summation of the partial DOS belonging to the TTF part also indicates the charge imbalance [Fig.~\ref{bands_P1bar}(d)]. 
It should be noted that GGA often underestimates the degree of charge disproportionation.  
However, here we would like to emphasize that even in GGA, a stable structure which breaks the $C$2/$c$ symmetry is found theoretically and shows a different electronic structure owing to the charge disproportionation between two types of molecules. 

\subsection{Quantum effects}
\begin{figure}
\begin{center}
\includegraphics[width=1.0\linewidth]{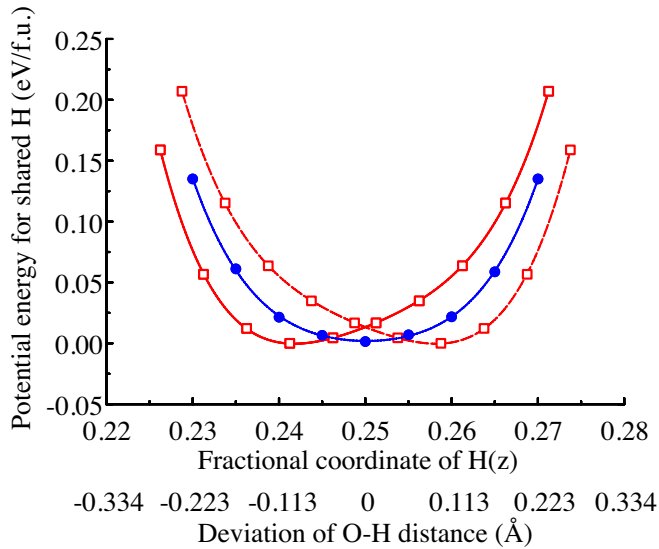}
\end{center}
\caption{(Color online) Potential energy curves for the shared H atom in $\kappa$-H$_3$(Cat-EDT-TTF)$_2$. 
Solid circles show potential energies for the optimized $C$2/$c$ structure. 
Open squares show potential energies for the $P\bar{1}$ structure. 
The origin of the longitudinal axis is set to be the minimum point of the $P\bar{1}$ structure.  
The horizontal axis is the fractional $z$~coordinate of the shared H atom located between two O atoms. 
H~($z$) = 0.25 corresponds to the center between two O atoms.}  
\setlength\abovecaptionskip{0pt}
\label{Hz_plot}
\end{figure}

Lastly, we discuss the quantum effects of the shared H atom. 
As shown above, we found a structure where the H atom locates close to an O atom, but there is a possibility that the quantum motion or tunneling of the proton makes it delocalized.~\cite{H_posi08} 
To estimate the degree of localization of the proton wave function, we plot the potential energy curve using the DFT total energies by changing the $z$~coordinate of the hydrogen position, where the positions of all the other atoms are fixed at the optimized $C$2/$c$ and $P\bar{1}$ structures, respectively~(Fig.~\ref{Hz_plot}). 
There are two curves in Fig.~\ref{Hz_plot} for the latter, due to the two equivalent local-minimum structure with H close to either O1 or O3. 
When the distance between the two minima is sufficiently short and the potential between the minima is shallow, 
the proton wavefunction becomes centered at the bond midpoint.~\cite{Cleland_Scie94_LBHB, LBHB_Science94, Dmarx_CPC05, HamadaMorikawaPRB10} 
This is actually the case here, and then we expect that the distribution of the proton can not be localized close to O1 or O3. 
The distribution should be the highest at the center of O1 and O3, and the H atom appears to be shared by the two O atoms. 
This is probably the situation in the actual material; the $C$2/$c$ structure is realized by such H delocalization in terms of the quantum motions of the protons. 
An interesting point is that the QSL state in $\kappa$-S is realized in the $C$2/$c$ structure where the dimers of the H(Cat-EDT-TTF) units are all crystallographically equivalent and can carry $S$~=~1/2 each. 
In clear contrast, in the $P\bar{1}$ structure showing the charge disproportionation, the existence of two kinds of dimers should break this situation.   
Therefore, the quantum effects may play a role for the fact that the localized spins remain down to the lowest temperature in $\kappa$-S.  

We also note that, in general, H-localized states with asymmetric O--H--O bonds have relatively larger curvature of the potential at the minimum~point, and their zero-point motion in terms of kinetic energy becomes higher~(lower in energy) than those of the symmetric bonds. However, in this system, the curvature at the minimum-point of the H-localized~($P\bar{1}$) structure is small and not markedly different from that of the $C$2/$c$ structure having a single minimum~(solid circles in Fig.~\ref{Hz_plot}). Therefore, the difference of zero-point energy should be small, resulting in a subtle situation where these two structures are energetically very competitive to each other. 

Then, by substituting a D atom for the shared H atom, it is natural to expect D to be localized around one of two O atoms due to the reduction of quantum fluctuation. In fact, deuterated samples of~$\kappa$-S were recently synthesized, and a phase transition to a D-localized state is observed.\cite{A_Ueda_DCat}   
The resultant expansion in the O$\cdots$O length is observed, which is known as the Ubbelohde effect.~\cite{Ubbelohde1939}  
Even in the D case, we expect that the relative stability between the $P\bar{1}$ and $C$2/$c$ structures is very subtle as well. Further studies, such as theoretical relaxations with hybrid functionals, are desired in the near future. 

\begin{table}
\caption{Theoretically optimized atomic positions for the $P\bar{1}$ structure of $\kappa$-H$_3$(Cat-EDT-TTF)$_2$. 
This structure has two inequivalent H(Cat- EDT-TTF) units, and the number of crystallographically independent atomic sites is twice as large as that of the $C$2/$c$ structure. 
The lattice vector of $P\bar{1}$ structure, $a_p$, $b_p$, $c_p$ is related to the conventional vector of the $C$2/$c$ structure, $a$, $b$, $c$, by $a_p$ = ($a - b$)/2, $b_p$ = ($a + b$)/2, $c_p$ = $c$. 
}
\label{P1bar_opt}
\begin{center}
\begin{tabular}{c}    
\begin{minipage}{0.5\hsize}
\begin{center}
\begin{tabular}{lccc}
\hline\hline
Label & \multicolumn{3}{c}{fractional coordinates} \\  
 & $x$ & $y$ & $z$ \\
\hline
O1 & 0.0117 & 0.0142 & 0.1506\\
O2 & 0.9879 & 0.9857 & 0.3462\\
O3 & 0.8338 & 0.2465 & 0.9825\\
O4 & 0.2437 & 0.8363 & 0.4819\\
H1 & 0.0096 & 0.9922 & 0.2412\\
H2 & 0.8922 & 0.1403 & 0.9502\\
H3 & 0.6170 & 0.3462 & 0.0801\\
H4 & 0.9867 & 0.1757 & 0.8292\\
H5 & 0.1743 & 0.9871 & 0.3287\\
H6 & 0.7252 & 0.4028 & 0.2511\\
H7 & 0.5814 & 0.5274 & 0.9831\\
H8 & 0.8414 & 0.4113 & 0.0263\\
H9 & 0.5971 & 0.2747 & 0.2491\\
H10& 0.5746 & 0.5051 & 0.7536\\
H11& 0.1368 & 0.8952 & 0.4485\\
H12& 0.5050 & 0.5747 & 0.2538\\
H13& 0.6539 & 0.3829 & 0.4200\\
H14& 0.5272 & 0.5816 & 0.4831\\
H15& 0.4103 & 0.8415 & 0.5263\\
C1 & 0.0135 & 0.1021 & 0.1657\\
C2 & 0.9220 & 0.2207 & 0.0786\\
C3 & 0.1012 & 0.0807 & 0.2636\\
C4 & 0.9142 & 0.3195 & 0.0934\\
C5 & 0.5299 & 0.3637 & 0.0706\\
C6 & 0.0804 & 0.1019 & 0.7639\\
C7 & 0.9003 & 0.9853 & 0.3337\\
C8 & 0.0948 & 0.1786 & 0.2769\\
\hline\hline
\end{tabular}
\end{center}
\end{minipage}

\begin{minipage}{0.5\hsize}
\begin{center}
\begin{tabular}{lccc}
\hline\hline
C9 & 0.9995 & 0.2995 & 0.1930\\
C10& 0.4906 & 0.7973 & 0.0122\\
C11& 0.1787 & 0.0948 & 0.7769\\
C12& 0.6155 & 0.5139 & 0.3045\\
C13& 0.5138 & 0.6156 & 0.8044\\
C14& 0.7810 & 0.0769 & 0.4214\\
C15& 0.6362 & 0.4701 & 0.4295\\
C16& 0.2998 & 0.9990 & 0.6930\\
C17& 0.3187 & 0.9143 & 0.5934\\
C18& 0.6105 & 0.6997 & 0.9316\\
C19& 0.1269 & 0.3209 & 0.3519\\
C20& 0.6997 & 0.6105 & 0.4318\\
C21& 0.3213 & 0.1265 & 0.8517\\
C22& 0.7973 & 0.4906 & 0.5123\\
C23& 0.6305 & 0.8325 & 0.0832\\
C24& 0.8321 & 0.6310 & 0.5834\\
S1 & 0.3555 & 0.8236 & 0.0165\\
S2 & 0.1602 & 0.2018 & 0.8955\\
S3 & 0.6695 & 0.5700 & 0.8079\\
S4 & 0.4675 & 0.9108 & 0.1268\\
S5 & 0.2025 & 0.1593 & 0.3955\\
S6 & 0.5703 & 0.6693 & 0.3079\\
S7 & 0.9947 & 0.4195 & 0.2202\\
S8 & 0.8237 & 0.3554 & 0.5165\\
S9 & 0.4200 & 0.9940 & 0.7204\\
S9 & 0.5800 & 0.0060 & 0.2796\\
S10& 0.7318 & 0.6942 & 0.9548\\
S11& 0.9105 & 0.4678 & 0.6271\\
S12& 0.6932 & 0.7327 & 0.4552\\
\hline\hline
\end{tabular}
        \end{center}
      \end{minipage}
    \end{tabular}
  \end{center}
\end{table}

\section{CONCLUSION}
We have investigated the structural and electronic properties of $\kappa$-H$_3$(Cat-EDT-TTF)$_2$ and $\kappa$-H$_3$(Cat-EDT-ST)$_2$ by first-principles DFT calculations. 
The calculated band structures show a quasi-2D character, but with a relatively large band dispersion along the interlayer direction compared to conventional molecular conductors having insulating layers. 
We also evaluated transfer integrals for the effective low-energy tight-binding model by fitting to the DFT bands, 
where the ratio of the interdimer transfer integrals representing the anisotropy of the triangular lattice, 
${b2}$/$p$ ($t^{\prime}$/$t$ in the literature), is 1.25 and 1.42 for $\kappa$-S and $\kappa$-Se, respectively. 
The character of the Kohn-Sham eigenstates of bands crossing the Fermi level is analyzed in detail for $\kappa$-S and found to be mainly constituted by the $p$~states of C and S atoms belonging to the TTF~part. 
As for the hydrogen bonds, we found a H-localized phase different from the experimental structure by structural optimization. 
It shows charge imbalance in the LDOS associated with two different H(Cat-EDT- TTF) units.
This phase may be related to the $P\bar{1}$ structure observed in deuterated $\kappa$-S system.
On the other hand, the distribution of the shared H atoms should be delocalized 
due to the quantum effects since the bottoms of the calculated potential wells are very shallow.
Our results suggest that the $C$2/$c$ structure of H$_3$(Cat-EDT-TTF)$_2$ can be stabilized 
by the quantum effects of H atoms. 

\section{ACKNOWLEDGEMENTS}
The authors thank T. Isono, A. Ueda, H. Mori~for stimulating discussion on the experimental aspect of the materials. 
We also thank K. Ueda, H. Otaki, S. Yamashita, M. Yamashita, and I. Hamada for fruitful discussions.
We acknowledge A. J. Freeman~for discussion about the FLAPW code. 
The computations in this work have been partly performed using the facilities of the
RIKEN Integrated Cluster of Clusters (RICC) and the numerical materials simulator
at the National Institute for Materials Science (NIMS), Japan. 
This work is supported by Grant-in-Aid for Scientific Research (Nos. 23108002, 26287070, 22224006, and 26400377) from MEXT~(Japan) and the RIKEN iTHES project. 
\newpage
\bibliography{./Tsumu}

\end{document}